\documentclass[aps,prfluids,preprint,groupedaddress]{revtex4-2}

\usepackage{amssymb,amsmath}
\usepackage{graphicx,placeins}

\begin{document}

\title{Nonlocality of the slip length operator for scalar and momentum transport in turbulent flow over superhydrophobic surfaces}

\author{Kimberly Liu}
\email[]{kimliu@stanford.edu}
\author{Ali Mani}
\email[]{alimani@stanford.edu}
\affiliation{Stanford University}

\date{\today}

\begin{abstract}
  Superhydrophobic surfaces (SHS) are textured hydrophobic surfaces which have the ability to trap air pockets when immersed in water. This can result in significant drag reduction, due to substantially lower viscosity of air resulting in substantial effective slip velocity at the interface. Past studies of both laminar and turbulent flows model this slip velocity in terms of a homogenized Navier slip boundary condition with a slip length relating the wall slip velocity to the wall-normal velocity gradient. In this work, we seek to understand the effects of superhydrophobic surfaces in the context of mean scalar and momentum mixing. We use the macroscopic forcing method (Mani and Park, 2021) to compute the generalized eddy viscosity and slip length operators of a turbulent channel over SHS, implemented as both a pattern-resolved boundary condition and homogenized slip length boundary condition, for several pattern sizes and geometries. We present key differences in the mixing behavior of both boundary conditions through quantification of their near-wall eddy viscosity. Analysis of transport in turbulent flow over pattern-resolved surfaces reveals substantial nonlocality in the measured homogenized slip length for both scalar and momentum mixing when the Reynolds and Peclet numbers based on pattern size are finite. We present several metrics to quantify this nonlocality and observe possible trends relating to Reynolds number, texture size, and pattern geometry. The importance of nonlocality in the slip length operator and in the eddy diffusivity operator is demonstrated by examining the impact on Reynolds-averaged solutions for the mean scalar and velocity fields.
\end{abstract}

\maketitle

\newpage

\section{\label{sec:intro}Introduction}
Superhydrophobic surfaces (SHS) are textured hydrophobic surfaces which have the ability to trap air pockets when immersed in water \cite{rothstein2010slip,golovin2016bioinspired}. This can result in significant drag reduction due to the substantial effective slip velocity which forms at the surface. Experimental studies of superhydrophobic surfaces in laminar flows have found drag reductions of more than $25\%$ \cite{ou2004laminar,ou2005direct,choi2006effective,choi2006large}. Extensions to turbulent flow show even further drag reduction capabilities, of up to $50\%$ \cite{daniello2009drag,woolford2009particle,park2014superhydrophobic}. Regarding practical scenarios in naval contexts, Gose et al. (2021) \cite{gose2021turbulent} have studied the performance of sprayed SHS with random roughness on a towed, submersible body, and reported drag reductions of $4-24\%$.

Past numerical studies of flows over SHS have focused on modeling mean field behavior utilizing direct numerical simulations \cite{min2004effects,martell2009direct}. Theoretical investigations of laminar flow have found that the effective boundary condition is a homogenized uniform slip length of the form

\begin{equation}
  \overline{u}_1\big|_{\text{wall}} = b_{\text{eff}} \frac{\partial \overline{u}_1}{\partial x_2}\bigg|_{\text{wall}}  
\end{equation}

\noindent where $\overline{u}_1|_{\text{wall}}$ is the averaged velocity on the boundary, $b_{\text{eff}}$ is the slip length, and $\frac{\partial \overline{u}_1}{\partial x_2}$ is the mean velocity gradient at the boundary \cite{philip1972flows,lauga2003effective}. In the low $Re$ limit, it can be shown that the slip length $b_{\text{eff}}$ depends solely on the texture geometry, including solid fraction $\phi_s$, and scales linearly with the surface texture size $\lambda$, but does not depend on the flow \cite{lauga2003effective,ybert2007achieving,belyaev2010effective,davis2010hydrodynamic}. This property is not preserved in the turbulent regime, where it has been found that $b_{\text{eff}}$ is not just a function of texture size, but may also depend on other parameters such as flow Reynolds number \cite{min2004effects,fukagata2006theoretical,martell2009direct,busse2012influence}. With regards to the bulk behavior, the magnitude of the fluid velocity profile is shifted by a slip velocity $u_s$. The buffer layer is shortened, and the log layer is shifted downwards by $\Delta U$ \cite{fukagata2006theoretical,busse2012influence}. For positive drag reduction, $\Delta U$ should be less than $u_s$. Importantly, the slope of the log layer is preserved. As such, the effects of a SHS on the mean velocity profile in the log layer can be modelled as

\begin{equation}
  \frac{\overline{u}_1(x_2)-u_s}{u_\tau} = \frac{1}{\kappa} \log \bigg(\frac{x_2 u_\tau}{\nu}\bigg) + B - \frac{\Delta U}{u_\tau}
  \label{eq:tbl}
\end{equation}

\begin{figure}[!htbp]
  \centering
  \includegraphics{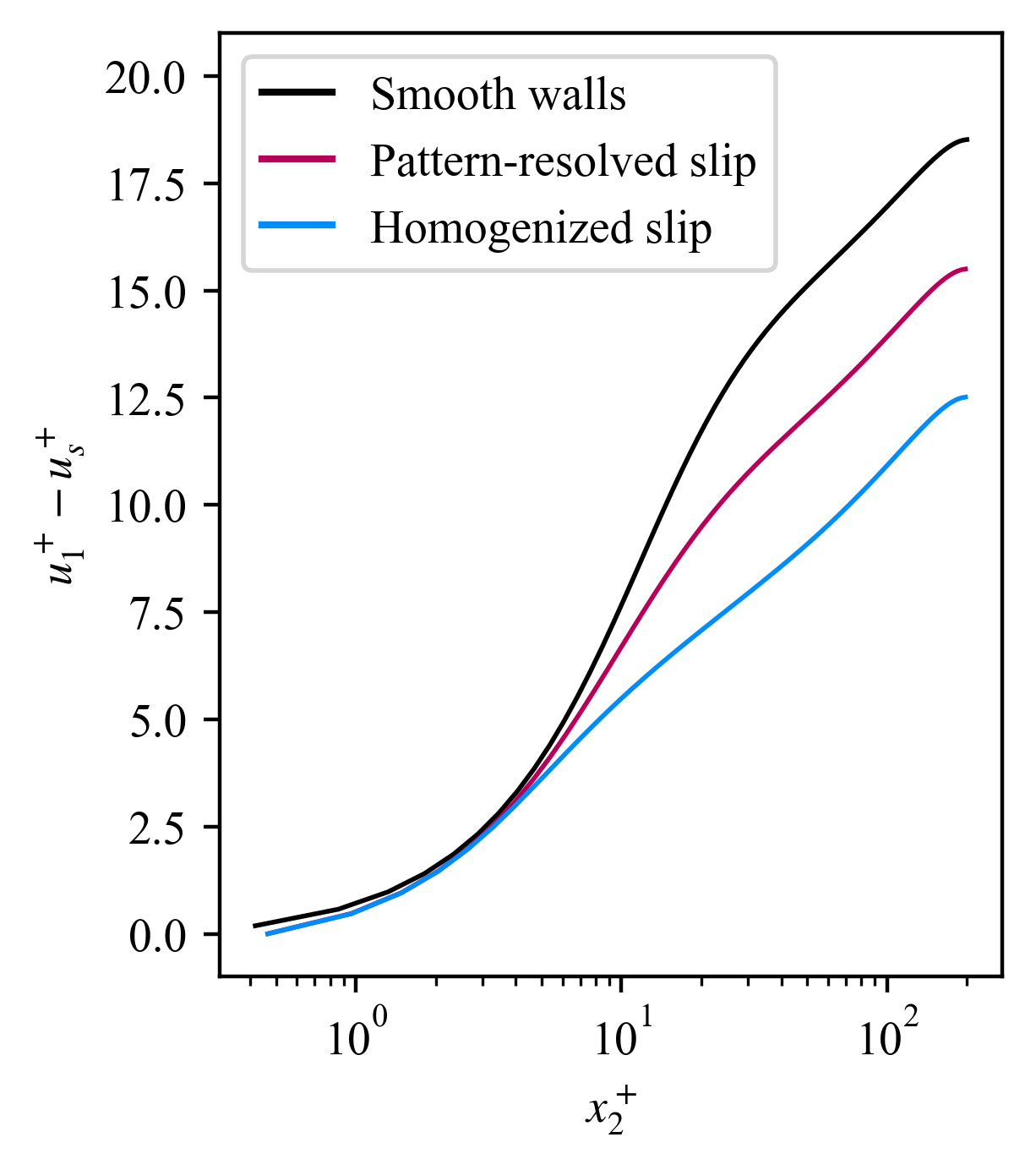}
  \caption{Mean streamwise velocity of smooth walls (black), pattern-resolved SHS (blue), and homogenized slip length SHS (red). The pattern-resolved SHS has texture size $\lambda^+ = 155.1$ and an observed slip velocity of $u_s^+ = 20.8$. Correspondingly, the homogenized SHS has slip length $b^+ = 20.8$.}
  \label{fig:mean_u}
\end{figure}

Direct numerical simulations (DNS) with a patterned slip boundary condition that resolve the pattern geometry have also been performed \cite{martell2009direct,martell2010analysis,park2013numerical,turk2014turbulent,jelly2014turbulence,lee2015effect,rastegari2015mechanism,jung2016effects,seo2016scaling}. Seo and Mani (2016) \cite{seo2016scaling} presented a scaling in the two asymptotic regimes of small $\lambda^+$ and large $\lambda^+$, where $\lambda^+ = \lambda u_\tau/\nu$, $u_\tau$ is the friction velocity, and $\nu$ is the kinematic viscosity of the fluid. Good agreement of wall shear and mean velocity profile between the homogenized slip and pattern-resolved slip boundary conditions was shown for small $\lambda^+$. For large texture sizes, a power law relationship was reported between $b^+$ and $\lambda^+$. Furthermore, it was found that DNS performed with the homogenized slip boundary condition does not match the mean flow of simulations with pattern-resolved slip for large $\lambda^+$ (Figure \ref{fig:mean_u}). This suggests that mean momentum mixing is influenced by the texture geometry for these flows.

In this work, we seek to understand the differences in the homogenized slip and pattern-resolved slip boundary conditions in the context of scalar and momentum mixing. We identify eddy diffusivity and eddy viscosity as quantitative indicators of scalar and momentum mixing, respectively. The macroscopic forcing method (MFM), a statistical technique that allows for direct measurement of the nonlocal eddy diffusivity operator from DNS data \cite{mani2021macroscopic}, is used to compare the eddy diffusivity and viscosity operators of a turbulent channel over patterned superhydrophobic surfaces, implemented as both a pattern-resolved boundary condition and corresponding homogenized slip length boundary condition. We also compare this data against eddy viscosities obtained from channel flow with solid walls, as studied by Park and Mani (2024) \cite{park2024direct}. We choose values of $\lambda^+ > \mathcal{O}(10)$ to investigate the limit of large $\lambda^+$, as presented by Seo and Mani (2016) \cite{seo2016scaling}.

We further extend the evaluation of nonlocal eddy viscosity to consideration of a nonlocal slip length operator boundary condition. Though the slip length boundary condition has, in the past, been limited to a dependence on local velocity gradients at the surface, we examine whether it can also depend on velocity gradients in a layer with finite thickness. This study is therefore a fundamental study using idealized configurations to provide a basic understanding of the nature of the slip mechanism for turbulent flows. In contrast to the mechanism in the Stokes limit, where the slip velocity solely depends on the velocity gradient on the wall, we demonstrate that, in the presence of turbulence, the slip is substantially influenced by the mean flow at a distance from the wall. 

\section{\label{sec:methods}Methods}
\subsection{Governing equations}

The incompressible Navier-Stokes equations are given below and consist of continuity and momentum transport.

\begin{subequations}
  \begin{eqnarray}
    \frac{\partial u_i}{\partial x_i} = 0 \\
    \frac{\partial u_i}{\partial t} + u_j \frac{\partial u_i}{\partial x_j} = -\frac{1}{\rho} \frac{\partial p}{\partial x_i} + \nu \frac{\partial^2 u_i}{\partial x_j \partial x_j}
  \end{eqnarray}
\end{subequations}

Here, $\rho$ is density, $p$ is pressure, and $\nu$ is kinematic viscosity. The scalar transport equations are

\begin{equation}
  \frac{\partial c}{\partial t} + u_j \frac{\partial c}{\partial x_j} = \alpha \frac{\partial^2 c}{\partial x_j \partial x_j} + S
\end{equation}

\noindent where $\alpha$ is diffusivity and S is a generic source term.

The Reynolds-averaged Navier-Stokes equations (RANS) are obtained by averaging the Navier-Stokes equations in homogeneous directions. In our problem, these are the streamwise ($x_1$) and spanwise ($x_3$) directions, as well as time.

\begin{equation}
  \frac{\partial \bar{u}_i}{\partial t} + u_j \frac{\partial \bar{u}_i}{\partial x_j} = -\frac{1}{\rho} \frac{\partial p}{\partial x_i} + \frac{\partial}{\partial x_j} \bigg( \nu \frac{\partial \bar{u}_i}{\partial x_j} \bigg) + \frac{\partial}{\partial x_j} \big( - \overline{u_j'u_i'} \big)
\end{equation}

The rightmost term is the divergence of Reynolds stress and is unclosed. It is often modelled with an eddy viscosity $D_v$, such that the original equation simplifies to

\begin{subequations}
  \label{eqn:RANS}
  \begin{eqnarray}
    -\overline{u_j'u_i'} = D_v \frac{\partial \bar{u}_i}{\partial x_j} \\
    \frac{\partial \bar{u}_i}{\partial t} + \bar{u}_j \frac{\partial \bar{u}_i}{\partial x_j} = -\frac{1}{\rho} \frac{\partial p}{\partial x_i} + \frac{\partial}{\partial x_j} \bigg( (\nu + D_v) \frac{\partial \bar{u}_i}{\partial x_j} \bigg)
  \end{eqnarray}
\end{subequations}

In the Reynolds-averaged scalar transport equation, the divergence of turbulent flux is modelled analogously, with eddy diffusivity $D_c$.

\begin{subequations}
  \label{eqn:RASTE}
  \begin{eqnarray}
    \frac{\partial \bar{c}}{\partial t} + \bar{u}_j \frac{\partial \bar{c}}{\partial x_j} = \frac{\partial}{\partial x_j} \bigg(\alpha \frac{\partial\bar{c}}{\partial x_j}\bigg) + \frac{\partial}{\partial x_j} \big(-\overline{u_j'c'}\big) + S \\
    -\overline{u_j'c'} = D_c \frac{\partial \bar{c}}{\partial x_j} \\
    \frac{\partial \bar{c}}{\partial t} + u_j \frac{\partial \bar{c}}{\partial x_j} = \frac{\partial}{\partial x_j} \bigg( (\alpha + D_c) \frac{\partial \bar{c}}{\partial x_j} \bigg) + S
  \end{eqnarray}
\end{subequations}

\subsection{Computational setup}

We perform direct numerical simulations of turbulent channel flow with homogenized slip and pattern-resolved boundary conditions. The three-dimensional Navier-Stokes and scalar transport equations are discretized and solved by code originally developed by Bose et al. (2010) \cite{bose2010grid} and modified for superhydrophobic textured boundaries by Seo and Mani (2016) \cite{seo2016scaling}. Simulations are performed at $Re_\tau \approx 197.5$ and $Pr = 1$. The computational domain is $2\pi h \times \pi h \times 2 h$ in the streamwise ($x_1$), spanwise ($x_3$), and wall-normal ($x_2$) directions, where $h$ is the channel half height. We use a second-order finite difference scheme on a staggered mesh, with uniform mesh spacing in the $x_1$ and $x_3$ directions, and stretched mesh in the $x_2$ direction. The grid resolution is $\text{max}(\Delta x_1^+) = 6.4$, $\text{max}(\Delta x_3^+) = 3.2$, $\text{min}(\Delta x_2^+) = 0.5$, and $\text{max}(\Delta x_2^+) = 6.4$. Additionally, grid resolution is chosen such that at least eight grid points resolve the texture width ($\Delta x_1^+ \leq w^+/8$ and $\Delta x_3^+ \leq w^+/8$). Second-order Adams-Bashforth and Crank-Nicolson schemes were used for time advancement \cite{kim1985application}.

\begin{figure}[!htbp]
  \centering
  \includegraphics[width=0.6\linewidth]{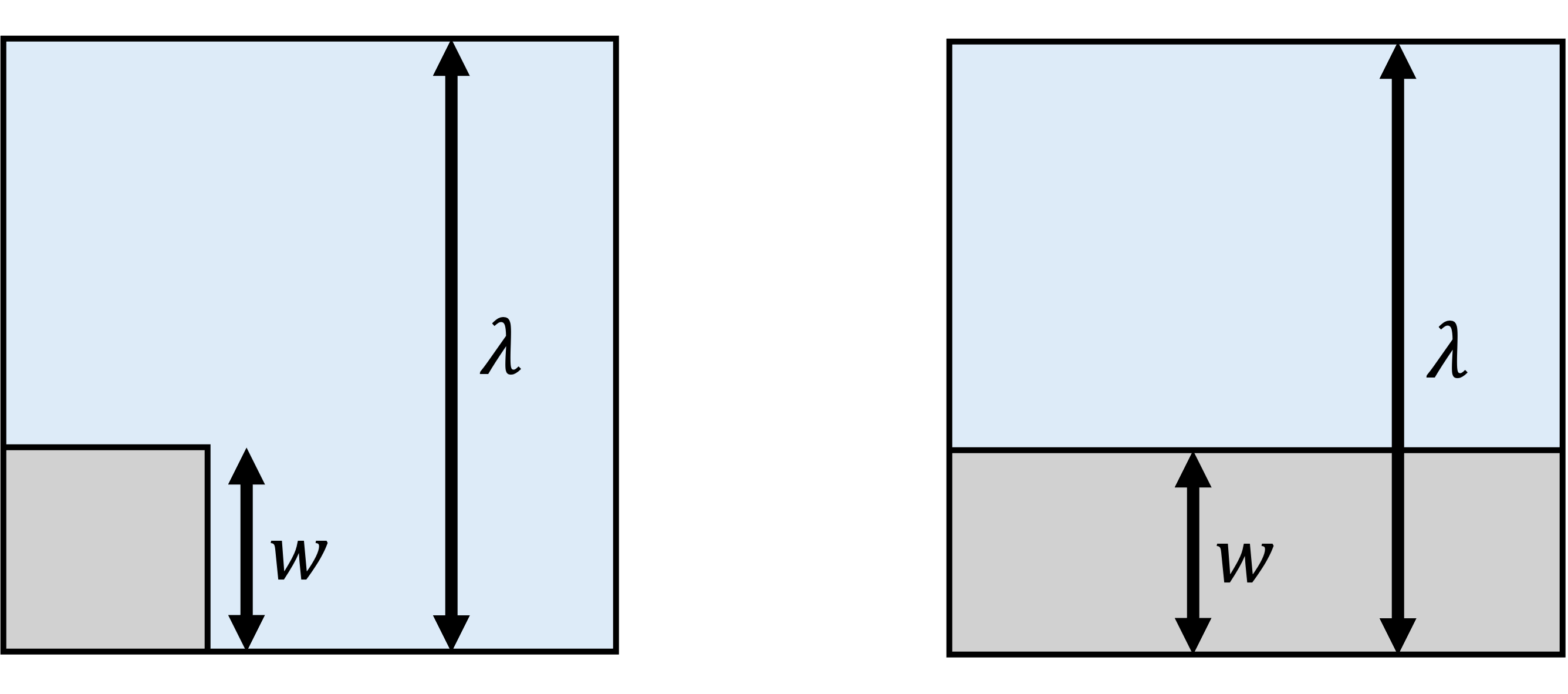}
  \caption{Schematic representation of a unit cell of superhydrophobic geometry: (a) square posts and (b) streamwise-aligned ridges. The streamwise flow is from left to right.}
  \label{fig:diagram}
  \end{figure}

For the gas-liquid interface, a Navier slip boundary condition was implemented for the homogenized slip cases and a patterned free-shear and no-slip condition was implemented for the pattern-resolved cases \cite{martell2009direct}. For thermal transport, these boundary conditions can be interpreted as adiabatic and isothermal boundary conditions, respectively. We assume infinite surface tension, zero Weber number, and flat interfaces for the pattern-resolved cases. Both the top and bottom wall are treated as superhydrophobic surfaces in all simulations. Simulations were conducted for square posts of texture wavelengths $\lambda^+ = 77.6$, $\lambda^+ = 103.4$, and $\lambda^+ = 155.1$, and corresponding homogenized slip lengths. In outer units, the listed pattern lengths correspond respectively to $\pi h /8$, $\pi h / 6$, and $\pi h / 4$. Streamwise-aligned ridges with texture wavelength of $\lambda^+=155.1$ were also simulated. All simulations were with solid fraction of $\phi_s = w^2/\lambda^2 = 1/9$. See Table \ref{tab:cases} for a summary of cases.

\begin{table}[!htbp]
\caption{Simulation cases\label{tab:cases}}
\begin{ruledtabular}
\begin{tabular}{|c|c|c|c|c|c|}
  Case & P077 & P103 & P155 & R155 & S155 \\
  \hline
  Geometry & Post & Post & Post & Ridge & Post \\
  \hline
  Transport & Momentum & Momentum & Momentum & Momentum & Scalar \\
  \hline
  $\lambda^+$ & 77.6 & 103.4 & 155.1 & 155.1 & 155.1 \\
\end{tabular}
\end{ruledtabular}
\end{table}

\begin{figure}[!htbp]
  \centering
  \includegraphics{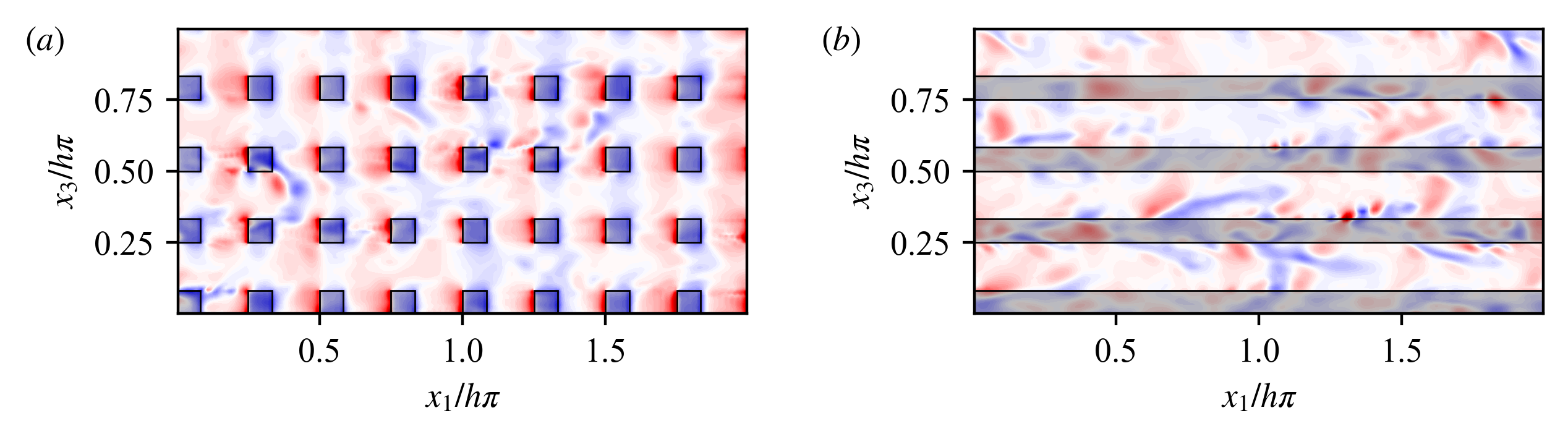}
  \caption{Instantaneous wall pressure of (a) square posts and (b) streamwise-aligned ridges with texture size $\lambda^+ = 155.1$.}
  \label{fig:pressure}
\end{figure}

\subsection{The macroscopic forcing method}

To evaluate eddy diffusivity and eddy viscosity operators, we utilize the macroscopic forcing method \cite{mani2021macroscopic}. Here, we shall present our methodology with the scalar transport equations, but this analysis is also applied to the momentum transport equations.

We apply forcing $s_\text{MFM}$ to the scalar field such that the mean scalar gradient takes on the Dirac delta ($\delta$) distribution.

\begin{equation}
  \frac{\partial c}{\partial t} + u_j \frac{\partial c}{\partial x_j} = \alpha \frac{\partial^2 c}{\partial x_j \partial x_j} + s_\text{MFM}
\end{equation}

\noindent Through careful selection of these forcings, one can measure the response of the turbulent flux to changes in the scalar gradient at other locations in the flow. The eddy diffusivity operator that we are interested in can be expressed in the general form \cite{berkowicz1979generalization,hamba2005nonlocal,kraichnan1987eddy}

\begin{equation}
  -\overline{u_j'c'}(\textbf{x}) = \int D_{c,jk}(\textbf{x},\textbf{y}) \frac{\partial \bar{c}}{\partial x_k} \bigg|_\textbf{y} d\textbf{y}
  \label{eq:scalar_visc}
\end{equation}

\noindent where $u_j'$ is the fluctuating part of the velocity field, $c'$ is the fluctuating part of the scalar field, $\bar{c}$ is the mean of the scalar field, $\textbf{x}$ is the location where the turbulent flux is measured, and $\textbf{y}$ is the location where the mean scalar gradient is measured.

In this form, the eddy diffusivity is a second-order tensorial kernel $D_{c,jk}$, which can account for both anisotropy and nonlocality. Since the DNS equation for scalar transport is linear, one expects that the linear projection of the system to macroscopic space would lead to a linear macroscopic operator. Eq. \ref{eq:scalar_visc} states the most general linear operator which relates mean concentration gradient to mean scalar flux for a statistically stationary problem. Writing the relation as a function of gradient of mean concentration, as opposed to the concentration itself, ensures that the flux can only be generated in the presence of mean scalar gradient; for a spatially uniform mean concentration, one expects zero fluctuations for an incompressible flow.

Specifically, we are interested in nonlocality of the eddy diffusivity kernel. Consider the problem in the context of the scalar flux transport equation \cite{younis2005rational}.  Due to history effects and other transport terms, a nonzero scalar flux can be created at one location and transported elsewhere in the domain. This would then lead to nonlocality in the implied eddy diffusivity kernel, once the kernel is projected to represent time-averaged behavior.

Using a Kramers-Moyal expansion \cite{van1992stochastic}, we can further decompose this eddy diffusivity into a leading-order term that characterizes local behavior and higher-order terms that characterize nonlocal behavior.

\begin{equation}
  -\overline{u_j'c'}(\textbf{x}) = 
  \bigg[ D_{c,jk}^0(\textbf{x}) + D_{c,jkm}^1(\textbf{x}) \frac{\partial}{\partial x_m} + \cdots \bigg] \frac{\partial \bar{c}}{\partial x_k}
\end{equation}

The leading-order coefficient, $D_{c,jk}^0(\textbf{x})$, is the zeroth moment of the eddy diffusivity kernel, and can be taken as a local approximation of the eddy diffusivity operator \cite{berkowicz1979generalization}:

\begin{subequations}
\begin{eqnarray}
  \label{eq:D0_def}
  D_{c,jk}^0(\textbf{x}) = \int D_{c,jk}(\textbf{x},\textbf{y}) d\textbf{y} \\
  -\overline{u_j'c'}(\textbf{x}) \approx D_{c,jk}^0(\textbf{x}) \frac{\partial \bar{c}}{\partial x_k}
\end{eqnarray}
\end{subequations}

This leading-order coefficient is identical to the full eddy diffusivity operator in the rare case that the kernel is a Dirac delta function and thus purely local. A counterexample of such a scenario would be Rayleigh-Bénard convection, which is highly nonlocal, where nonzero mean turbulent scalar flux can be measured in a zone of zero mean scalar gradient \cite{shang2004measurements,shishkina2007local}. As we shall see for transport on superhydrophobic surfaces, similar nonlocalities both for scalar and momentum transport can be quantified.

Because we are simulating a turbulent channel, we will focus on the components corresponding to the dominant scalar gradient $\frac{\partial\bar{c}}{\partial x_2}$. Thus,

\begin{equation}
  -\overline{u_2'c'}(x_2) = \int D_{c,22}(x_2,y_2) \frac{\partial \bar{c}}{\partial x_2} \bigg|_{y_2} dy_2 
\end{equation}

And the leading order approximation is

\begin{equation}
  -\overline{u_2'c'}(x_2) \approx D_{c,22}^0(x_2) \frac{\partial \bar{c}}{\partial x_2} \bigg|_{x_2}
\end{equation}

Similar analysis is carried out for the momentum transport equation. The generalized eddy viscosity operator and relevant leading order approximation are

\begin{subequations}
\begin{eqnarray}
  -\overline{u_i'u_j'}(\textbf{x}) = \int D_{v,ijkl}(\textbf{x},\textbf{y}) \frac{\partial \bar{u}_l}{\partial x_k} \bigg|_\textbf{y} d\textbf{y} \label{eqn:gen_eddy_visc} \\
  -\overline{u_2'u_1'}(x_2) \approx D_{v,2121}^0(x_2) \frac{\partial \bar{u}_1}{\partial x_2} \bigg|_{x_2}
\end{eqnarray}
\end{subequations}

\subsection{Extension of MFM to slip length operator}

We further extend the macroscopic forcing method to evaluate a generalized slip kernel of the form

\begin{subequations}
  \begin{eqnarray}
    \bar{c}\big|_\text{wall} = \int \beta_c (y_2) \frac{\partial \bar{c}}{\partial x_2} \bigg|_{y_2} dy_2 \\
    \bar{u}_1\big|_\text{wall} = \int \beta_v (y_2) \frac{\partial \bar{u}_1}{\partial x_2} \bigg|_{y_2} dy_2
  \end{eqnarray}
\end{subequations}

\noindent where $\beta(y_2)$ is the generalized slip kernel. This form allows for nonlocal effects of the scalar or velocity gradient on the slip boundary condition at the wall. As with the eddy diffusivity and eddy viscosity, we can identify a leading order moment for the slip kernel.

\begin{subequations}
\begin{eqnarray}
  \label{eq:B0_def}
  \beta^0 = \int \beta(y_2) dy_2 \\
  \bar{c}\big|_\text{wall} \approx \beta_c^0 \frac{\partial \bar{c}}{\partial x_2} \\
  \bar{u}_1\big|_\text{wall} \approx \beta_v^0 \frac{\partial \bar{u}_1}{\partial x_2}
\end{eqnarray}
\end{subequations}

In the Stokes flow limit ($Re \ll 1$), the velocity slip length kernel simplifies to a $\delta$ function, corresponding to purely local behavior as represented by the standard homogenized slip length boundary condition \cite{lauga2003effective,ybert2007achieving} and by the leading order moment. We will evaluate the kernel at finite $Re$ to quantify any possible nonlocality that arises.

At this point, we would like to note that previous MFM analysis has only been performed on flows where the boundary conditions are already consistent with macroscopic forcing. The question arises of how to define a macroscopic boundary condition for our forced equations that is consistent with the pattern-resolved slip boundary condition in the DNS equations. We relax the no-slip boundary condition on the solid surfaces of the pattern geometry to allow for a mean, forced value $s_c$ or $s_v$ for scalars and momentum, respectively.

\begin{subequations}
  \begin{eqnarray}
    c_\text{solid} = s_c \\
    \frac{\partial c_\text{air}}{\partial x_2} = 0
  \end{eqnarray}
\end{subequations}

\noindent This forced value is chosen such that the mean over the total patterned surface is compatible with the macroscopic forcing, and is recalculated at each time-step.

\section{\label{sec:scalars}Scalar Transport}
\subsection{\label{subsec:scalar_intro}Problem formulation}

To motivate our consideration of nonlocal effects on slip length, we shall first consider the problem of scalar mixing. The scalar transport equations are linear, and so present a simplified baseline from which to analyze nonlocality. For this problem, we shall choose temperature as our scalar, but the results shown here are generally applicable to any passive scalar. Past work has shown that patterned slip surfaces increase scalar transport of many interfacially driven transport phenomena \cite{haase2013momentum,ajdari2006giant,bocquet2007flow}.

\begin{equation}
  \frac{\partial T}{\partial t} + u_j \frac{\partial T}{\partial x_j} = \alpha \frac{\partial^2 T}{\partial x_j \partial x_j} + \frac{\dot{q}''}{\rho c_p}
\end{equation}

\noindent Here, $\dot{q}''$ is volumetric heat generation and $c_p$ is specific heat. For convenience, we shall use $Q = \frac{\dot{q}''}{\rho c_p}$.

Consider the problem of turbulent channel flow. Both top and bottom walls of the channel are superhydrophobic surfaces with square post geometry, of solid fraction $\phi_s = 1/9$ and texture wavelength $\lambda^+ = 155.1$. The bottom wall is held at a temperature of $+T_\text{wall}$ and the top wall is held at a temperature of $-T_\text{wall}$. For simplicity, we choose $T_\text{wall} = 1$. We will compare the scalar mixing of cases with and without volumetric heating $Q$. 

We can fit a slip length for this problem formulation as below. For temperature, this slip length has also been defined as the Kapitsa length \cite{bocquet2007flow,swartz1989thermal,patel2005thermal}.
\begin{equation}
  b_T^+ = \frac{T-T\big|_\text{wall}}{\partial T/\partial y^+}
\end{equation}

In the low thermal Peclet limit, slip length is shown to solely depend on the pattern geometry \cite{lauga2003effective}; in the finite thermal Peclet limit, it is expected to additionally depend on the underlying flow and molecular diffusivity \cite{barrat2003kapitza,patel2005thermal}. But it should not be affected by the presence of uniform heat generation. To test this premise, we choose $Q = 0.025$, which produces a mean temperature profile in which $\frac{\partial T}{\partial x_2}$ varies significantly in the near-wall region of the top wall.

\begin{figure}[!htbp]
  \centering
  \includegraphics{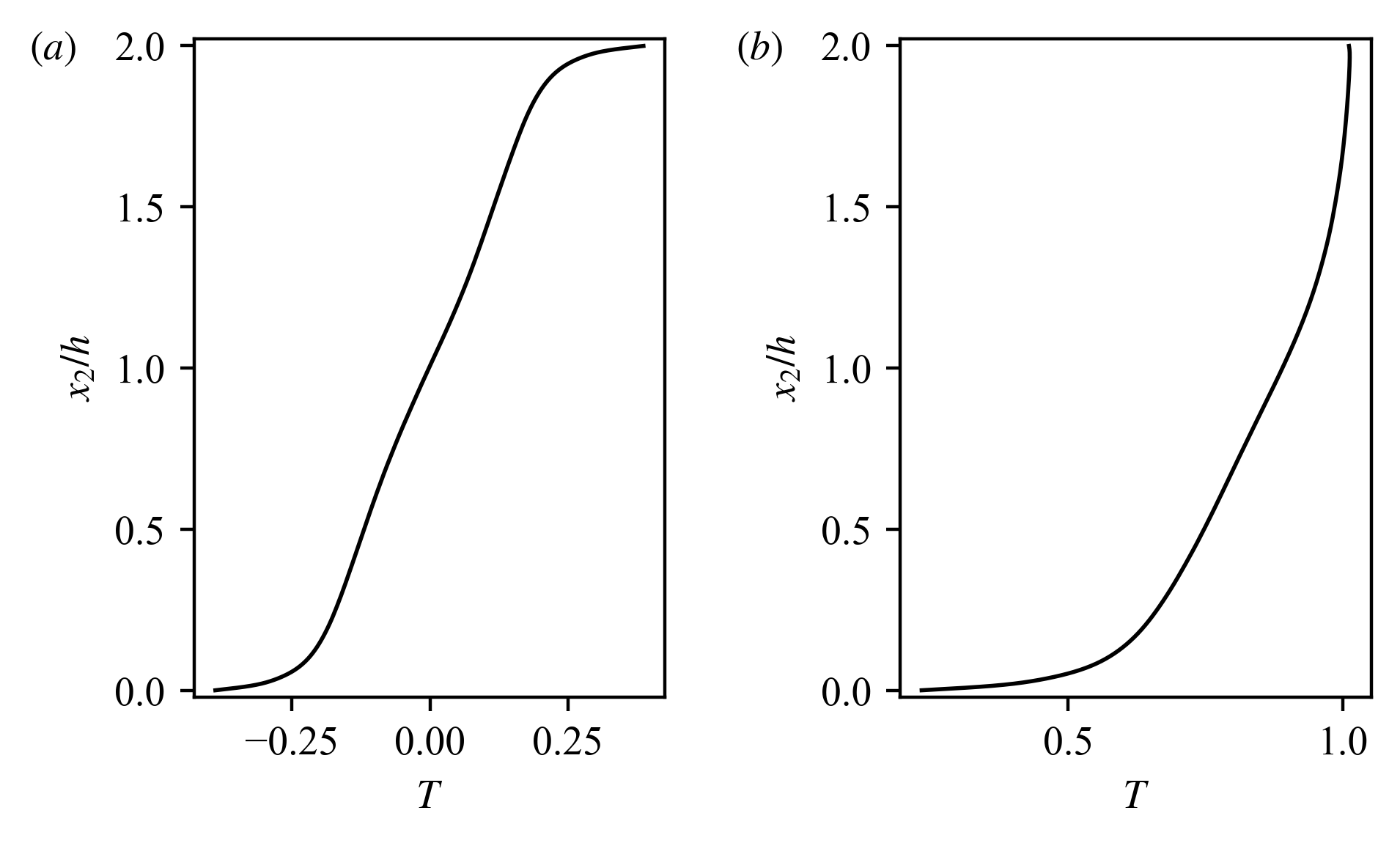}
  \caption{Mean temperature profiles from DNS of (a) $Q = 0$ and (b) $Q \neq 0$}
  \label{fig:scalar_profs}
\end{figure}

We can then calculate the slip length for each of these cases. The average slip length of the zero $Q$ case is $b^+ = 23.3$. For the nonzero $Q$ case, we will consider the slip length of the bottom and top walls separately. The bottom wall has an average slip length of $b^+ = 23.2$ and a negligible deviation of $0.16\%$ from the fitted value based on the zero heating case. The top wall, which is where we see significant variance in temperature gradient, has average slip length $b^+= 21.5$ and a substantial deviation of $8\%$.

This is a remarkable result. Two walls, from the same simulation and with the same boundary conditions, exposed to statistically identical flows, have markedly different slip lengths. In other words, the local slip length formulation is not robust to variations in mean temperature gradient. Thus, we are motivated to utilize MFM to measure the effects of nonlocal gradients. In the MFM equations, the heating term is absorbed into the forcing, and so we can measure a slip operator which is independent of any volumetric heating.

\subsection{Nonlocal operators for scalar transport}

\begin{figure}[!htbp]
  \centering
  \includegraphics{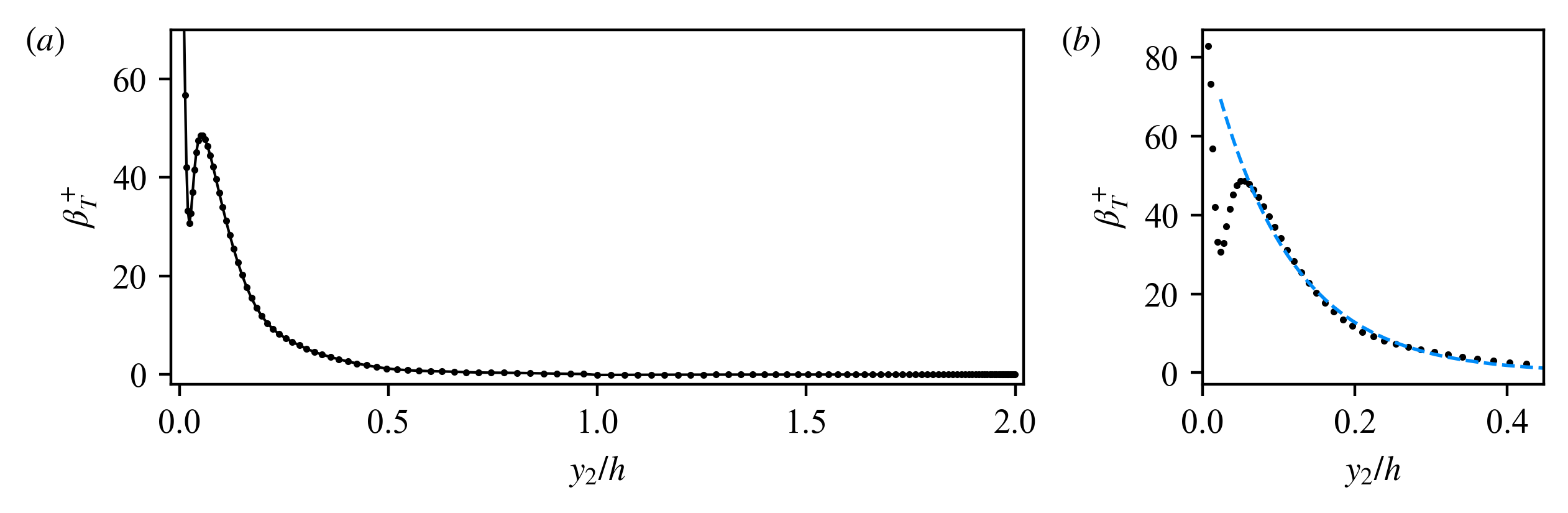}
  \caption{(a) Nonlocal slip kernel of scalar transport for pattern geometry $\lambda^+ = 155.1$. The $\delta$ function contributions close to the wall are not shown, for readability. (b) Exponential fit to nonlocal slip kernel}
  \label{fig:scalar_slip_kernel}
\end{figure}

Figure \ref{fig:scalar_slip_kernel}a shows the generalized slip length kernel as calculated by MFM. We observe that this kernel can be approximately represented as a $\delta$ function at $x_2 = 0$ and a decaying exponential that represents nonlocal effects of the SHS patterning. From this kernel, we proceed to quantify nonlocality in two ways. First, by fitting a decaying exponential to the slip length kernel (Figure \ref{fig:scalar_slip_kernel}b), we can define a length scale $\ell^+$ that corresponds to the rate of decay.

\begin{equation}
    \beta_T^+(y_2^+) \approx a \cdot \exp \bigg( -\frac{y_2^+}{\ell^+} \bigg)
\end{equation}

The decay length scale is $\ell^+ = 20.6$. We shall later compare this value to results from varying pattern geometries and momentum transport (Table \ref{tab:decay}).

The second method by which we propose to quantify nonlocality is to integrate over the slip length kernel and determine the percentage of the kernel that is influenced by local $y_2$. That is, how much the nonlocal velocity gradients contribute to the slip velocity, compared to the contributions of local velocity gradient at the wall. For this case, the local contributions are $\int_0^{0^+} \beta^+_T dy_2 = 22.16$, the nonlocal contributions are $\int_{0^+}^{2h} \beta_T^+ dy_2 = 9.01$, and the total integral is $\int_0^{2h} \beta^+_T dy_2 = 31.17$, for a nonlocality percentage of $29\%$ (Table \ref{tab:nonlocal}).

We note that neither the local contributions of the slip kernel nor the total contributions of the slip kernel are equal to the measured slip length of the zero heating case, $23.3$. In Section \ref{subsec:scalar_intro}, we found that this slip length is not generalizable to nonzero heating. In contrast, the slip kernel presented here is a unified slip length model which is applicable to both zero and nonzero heating, and should give consistent results on both the top and bottom walls. We will presently investigate the performance of this nonlocal slip kernel in Section \ref{subsec:RASTE}.

\begin{figure}[!htbp]
  \centering
  \includegraphics{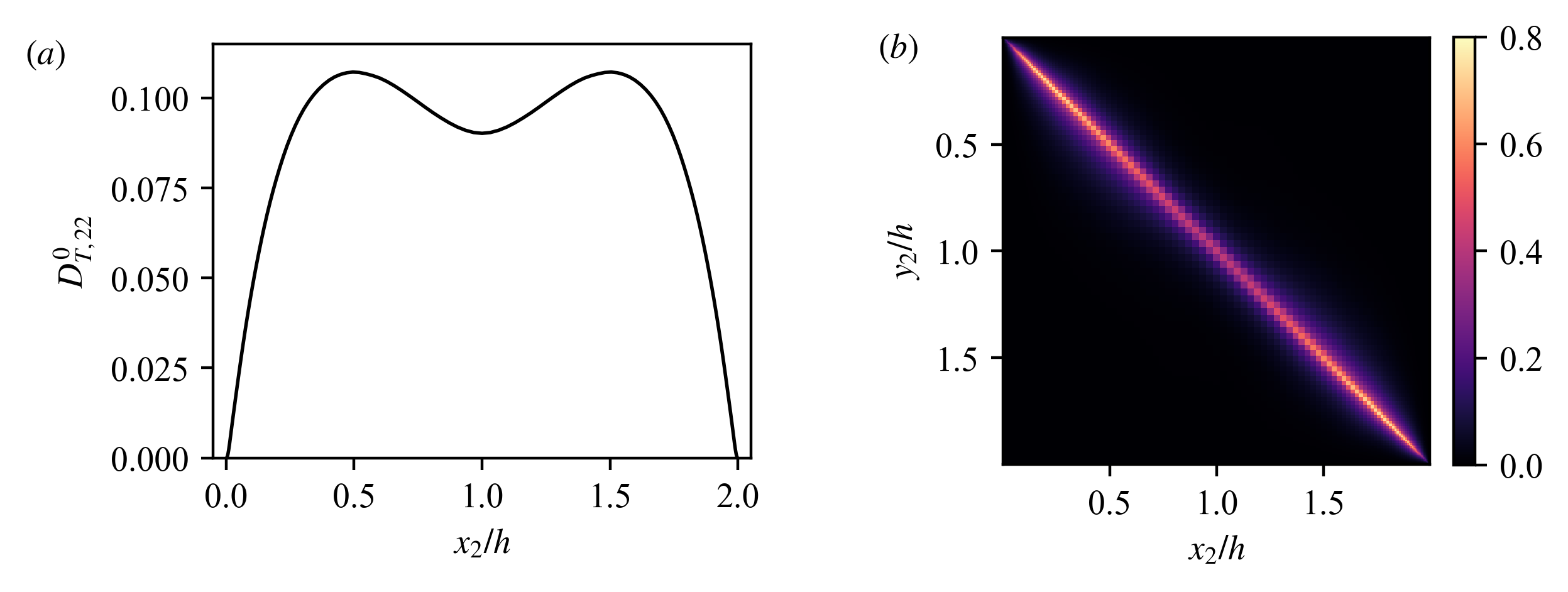}
  \caption{(a) Leading order eddy diffusivity $D_{T,22}^0$ and (b) nonlocal eddy diffusivity kernel $D_{T,22}$ for scalars with patterning $\lambda^+ = 155.1$.}
  \label{fig:scalar_nonlocal_diffusivity}
\end{figure}

\begin{figure}[!htbp]
  \centering
  \includegraphics{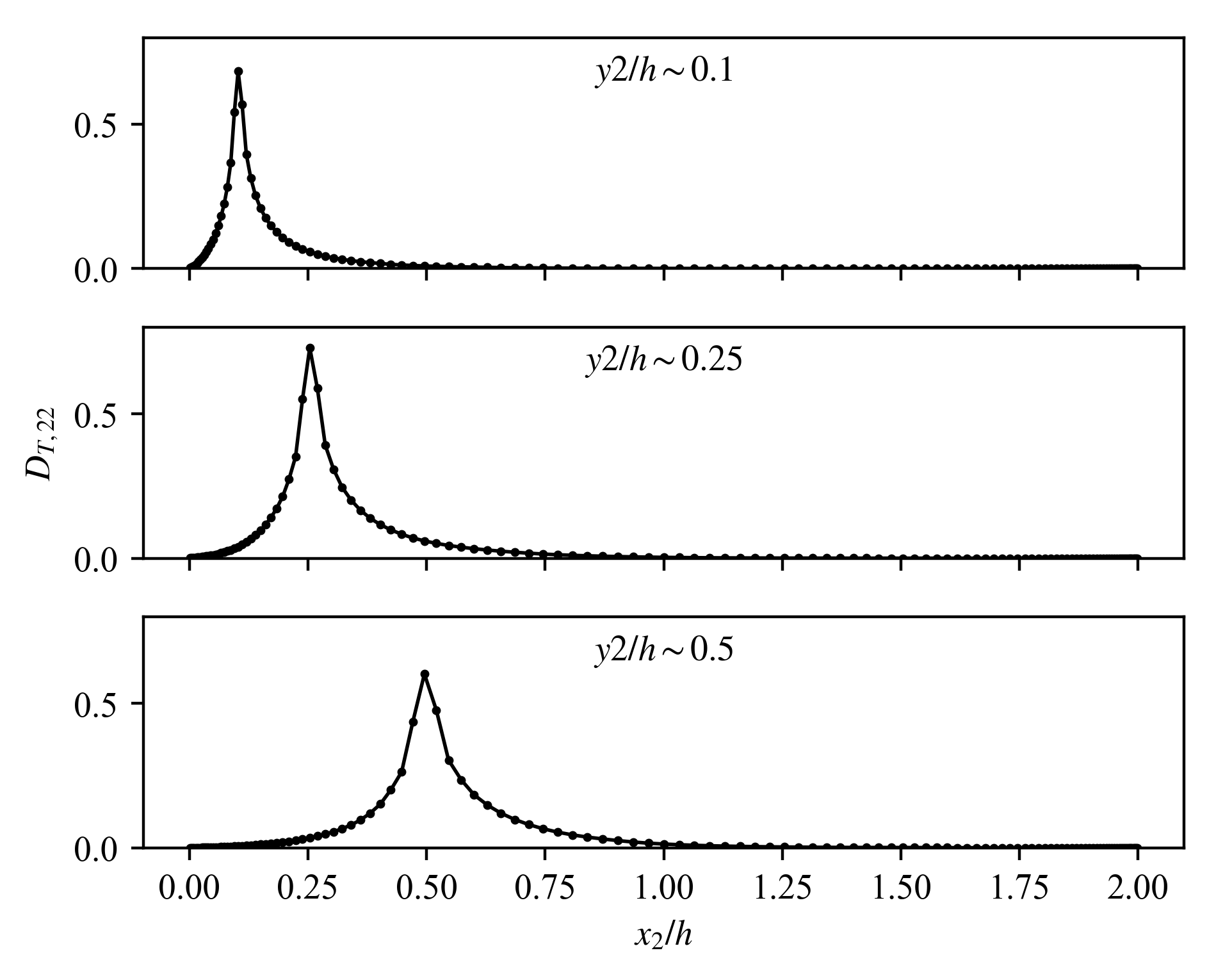}
  \caption{Profiles of nonlocal eddy diffusivity kernel $D_{T,22}$ for fixed $y_2$.}
  \label{fig:scalar_D22_cuts}
\end{figure}

Figure \ref{fig:scalar_nonlocal_diffusivity}a shows the leading order eddy diffusivity and Figure \ref{fig:scalar_nonlocal_diffusivity}b shows the full eddy diffusivity kernel for $\lambda^+ = 155.1$. A purely local eddy diffusivity kernel would only be active on the diagonal, where $x_2$ and $y_2$ values are equal. In contrast, there is significant nonlocality present in our patterned SHS case, represented by the spread in nonzero values of the kernel (Figure \ref{fig:scalar_D22_cuts}).

\subsection{Reynolds-averaged scalar transport assessment}
\label{subsec:RASTE}

To assess the impact of the measured nonlocalities, we shall utilize the eddy diffusivity and slip length operators, as calculated by MFM, to solve the Reynolds-averaged scalar transport equations (Eq. \ref{eqn:RASTE}). Solutions were calculated with leading order eddy diffusivity $D_{T,22}^0$ and either local slip length $\beta_T^0$ or nonlocal slip kernel $\beta_T(y_2)$. Note that, because we utilize the same eddy diffusivity operator for both solutions, the difference in local and nonlocal slip will only affect the result at the boundary. We would like to emphasize that the same eddy diffusivity and slip length operators are used for both the zero and nonzero heating cases.

For both cases, using only the leading order eddy diffusivity and local slip length results in large error in all regions of the domain. The maximum error between the DNS and Reynolds-averaged solutions is a difference of $0.07$ for the zero heating case and $0.26$ for the nonzero heating case, both at the boundaries. The introduction of a nonlocal slip length kernel shows dramatic improvement in the near wall region. The maximum error is $0.04$ for the zero heating case and $0.08$ for the nonzero heating. When using the nonlocal slip kernel, the error in the Reynolds-averaged solution of the nonzero heating case is now on the same order as that of zero heating.

In the zero heating case, the RMS errors are $0.03$ for the solutions with local operators and $0.01$ for the solutions with nonlocal operators. In the nonzero heating case, the RMS errors are $0.16$ for the solutions with local operators and $0.02$ for the solutions with nonlocal operators.

\begin{figure}[!htbp]
  \centering
  \includegraphics{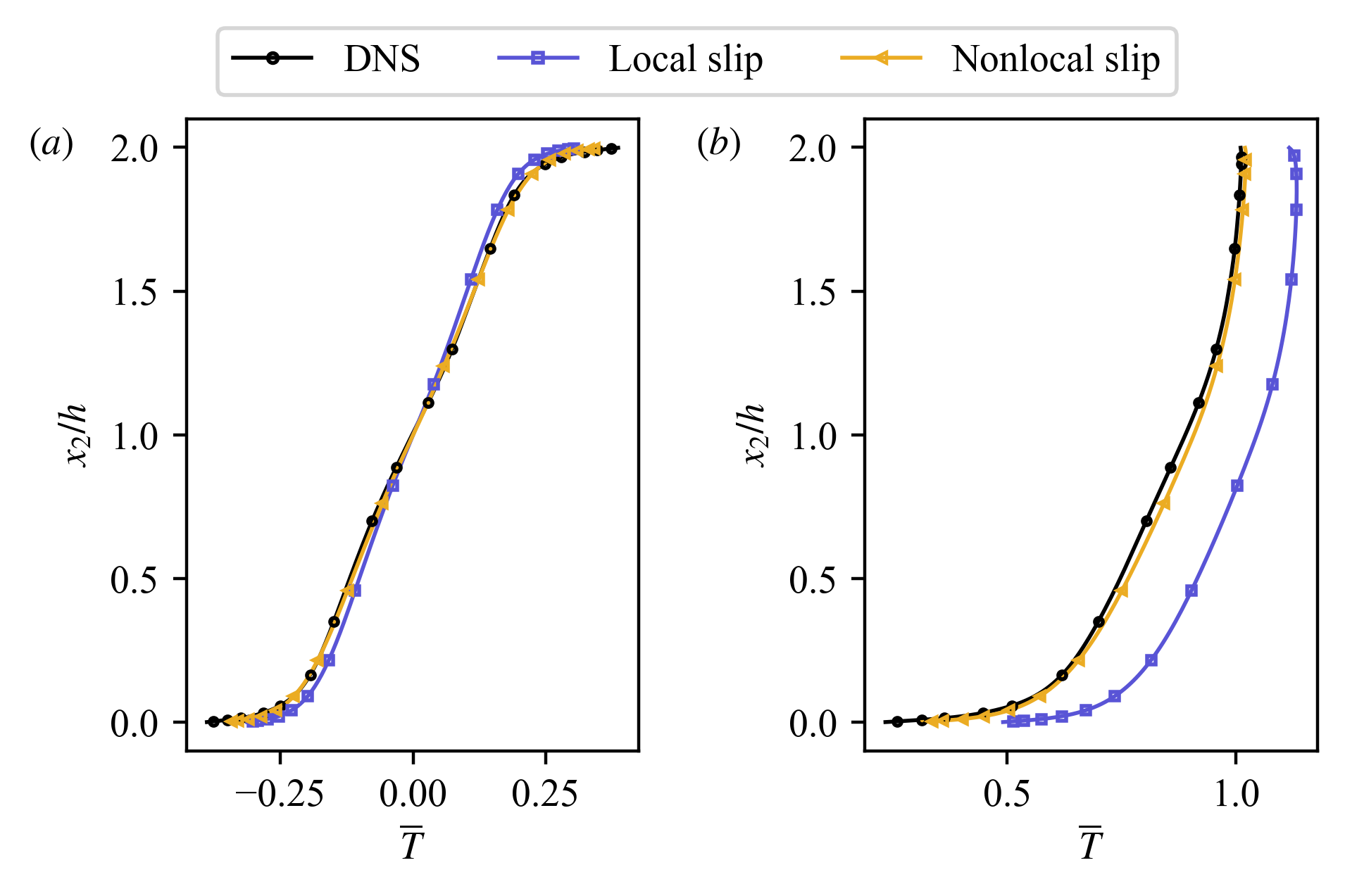}
  \caption{Mean temperature profiles for (a) $Q = 0$ and (b) $Q \neq 0$ from DNS (black), Reynolds-averaged solutions with local slip (purple), and Reynolds-averaged solutions with nonlocal slip (yellow).}
  \label{fig:scalar_rans}
\end{figure}

\section{\label{sec:results}Momentum Transport}
\subsection{Comparison of homogenized slip and pattern-resolved slip}

\begin{figure}[!htbp]
  \centering
  \includegraphics{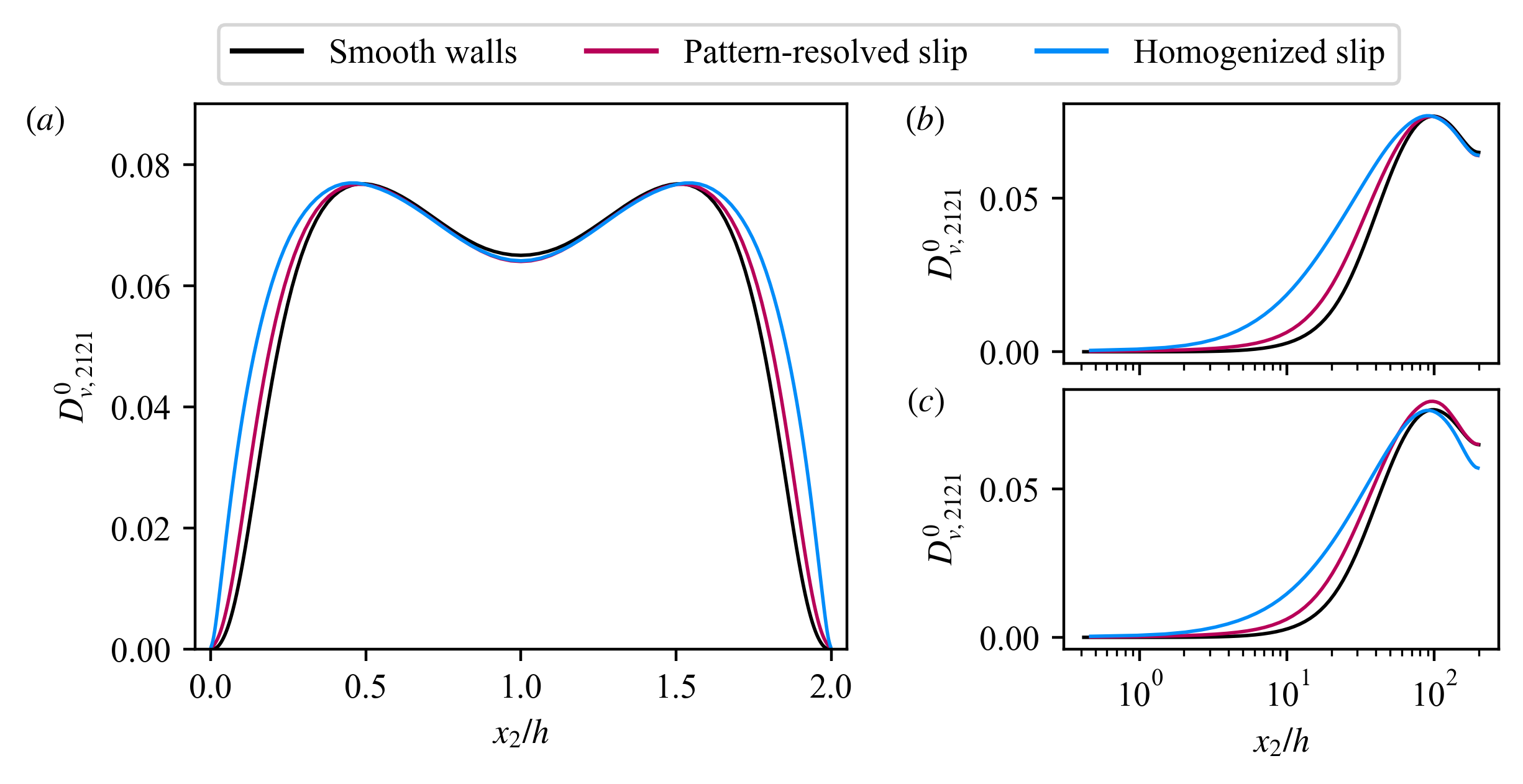}
  \caption{Leading order eddy viscosity component $D_{v,2121}^0$ of (a) $\lambda^+ = 155.1$, (b) $\lambda^+ = 155.1$ on log scale, and (c) $\lambda^+ = 77.6$ on log scale. Eddy viscosity was computed from MFM for smooth walls (black), pattern-resolved SHS of square posts (blue), and homogenized slip length SHS (red).}
  \label{fig:D0_comps}
\end{figure}

Our measurements of leading order eddy viscosity $D_{v,2121}^0$ for square posts of $\lambda^+ = 155.1$ and the corresponding $b^+_\text{eff} = 20.8$ reveal that the eddy viscosity operator for SHS is distinctly different from that of a smooth wall (Figure \ref{fig:D0_comps}), particularly in the inner zone of the flow. Specifically, the eddy viscosity for both pattern-resolved and homogenized slip SHS in this zone is larger in magnitude than that of smooth walls, indicating that high mixing is pushed closer to the wall due to the slip behavior near the surface. This enhanced mixing in the inner zone is consistent with observed extension of the log-layer towards the wall and thus shortening of the buffer layer leading to positive $\Delta U$ (momentum deficit) in the mean velocity profile.

Moreover, our results show that pattern-resolved SHS has substantially higher mixing, i.e., higher eddy viscosity, near the wall, with considerable momentum mixing pushed towards the viscous sublayer ($x_2^+<5$) when  compared to the homogenized slip SHS. This observation is in spite of the fact that the pattern-resolved system has an equivalent effective slip length to the homogenized slip case. While we expect that the usage of a homogenized slip length model in place of the more computationally expensive pattern-resolved SHS is not well-founded for our largest choice of $\lambda^+$, we find that this observation remains true for smaller $\lambda^+$ as well (Figure \ref{fig:D0_comps}c). The significant difference in mixing behavior close to the wall may indicate that the presence of the pattern geometry is critical for all patterned SHS, regardless of the magnitude of $\lambda^+$.

\subsection{Nonlocal operators for momentum transport}

\begin{figure}[!htbp]
  \centering
  \includegraphics{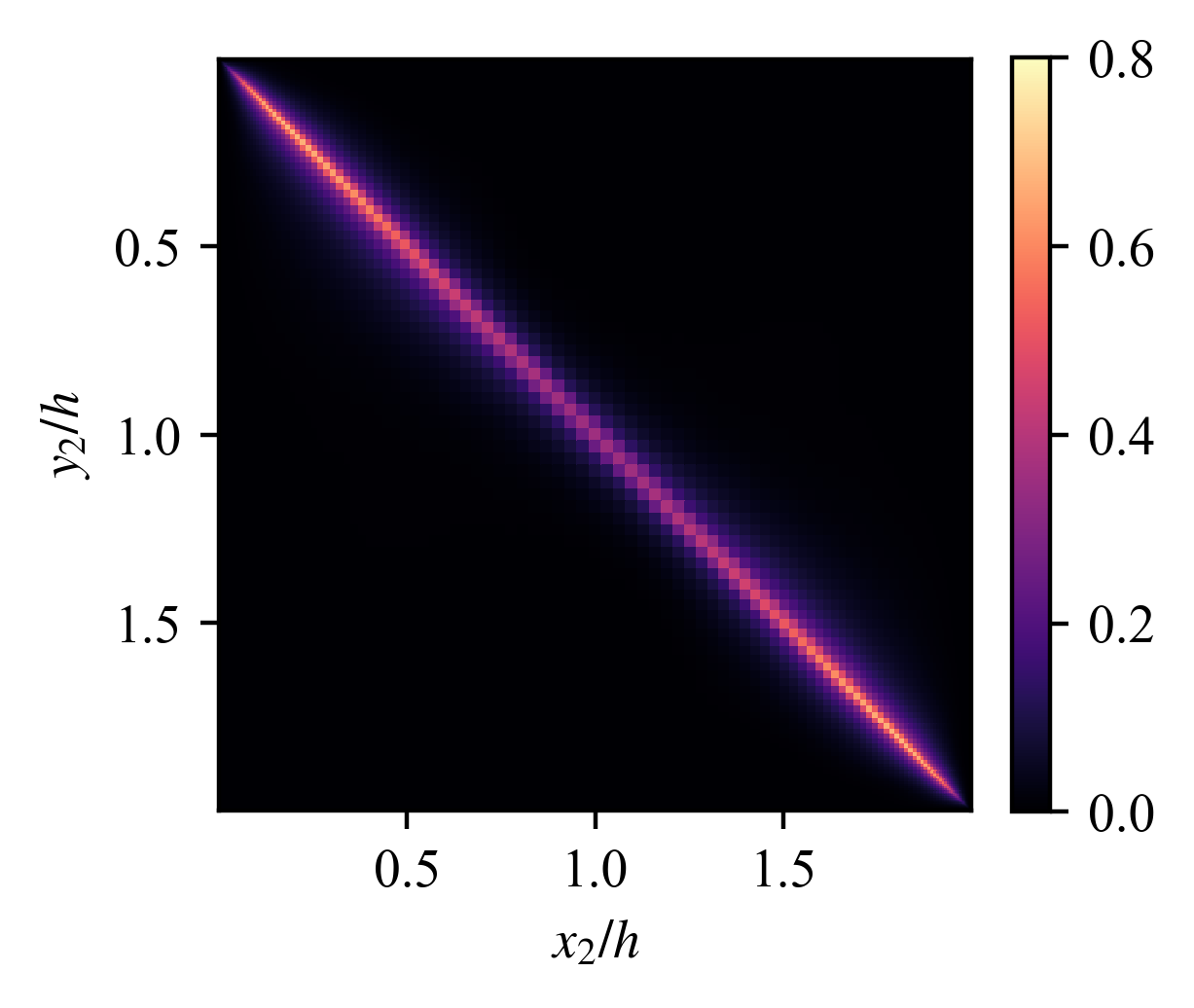}
  \caption{Nonlocal eddy viscosity kernel $D_{v,2121}(x_2,y_2)$ for $\lambda^+ = 155.1$}
  \label{fig:D2121}
\end{figure}

\begin{figure}[!htbp]
  \centering
  \includegraphics{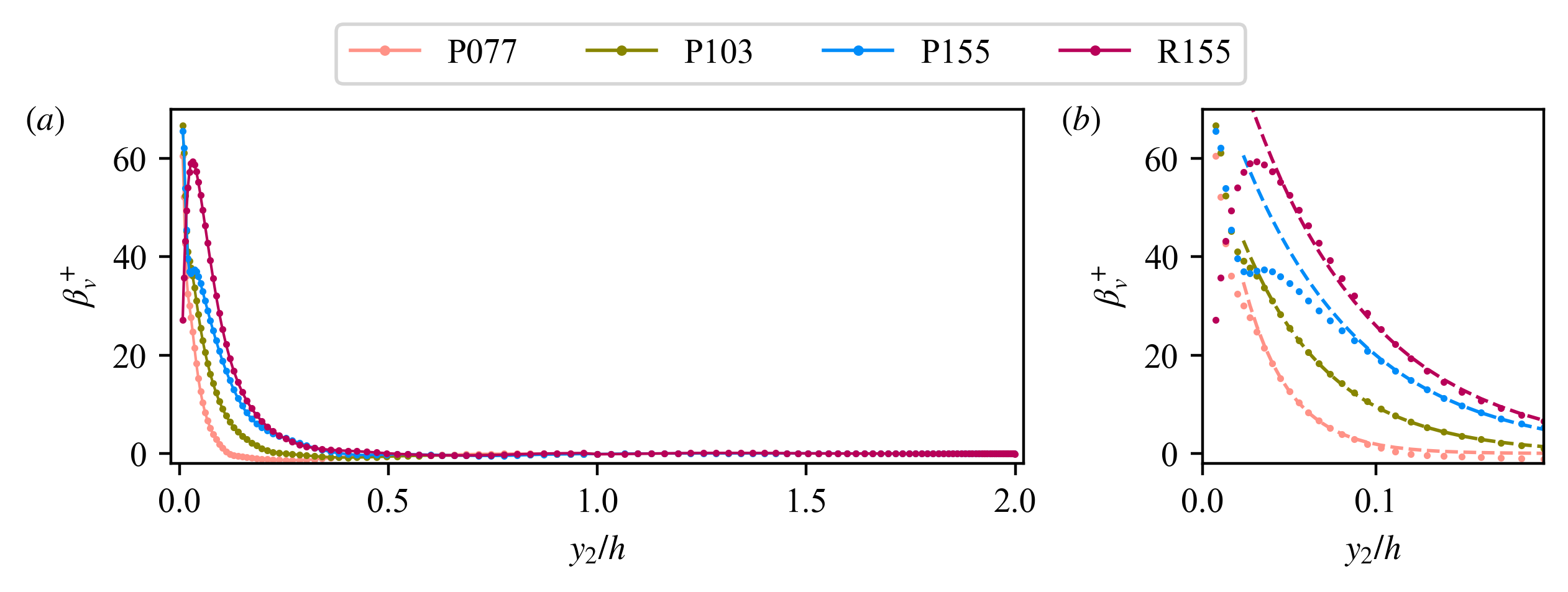}
  \caption{(a) Nonlocal slip kernel of momentum transport for $\lambda^+ = 77.6$ (orange), $\lambda^+ = 103.4$ (green), and $\lambda^+ = 155.1$ (blue) for square posts and $\lambda^+ = 155.1$ (red) for streamwise-aligned ridges. The $\delta$ function contributions close to the wall are not shown, for readability. (b) Exponential fits to nonlocal slip kernel}
  \label{fig:slip_kernel}
\end{figure}

Figure \ref{fig:D2121} shows the full eddy viscosity kernel for $\lambda^+ = 155.1$. As with the eddy diffusivity kernel for scalar transport, there is significant nonlocality. The generalized slip length kernel is plotted in Figure \ref{fig:slip_kernel}a for square posts of $\lambda^+ = 77.6$, $\lambda^+ = 103.4$, and $\lambda^+ = 155.1$, and streamwise-aligned ridges of $\lambda^+ = 155.1$. As with eddy viscosity, the trends of the slip length kernel for momentum reflect what we previously observed with scalar transport. We once more observe a local $\delta$ function and a nonlocal effect that resembles a decaying exponential.

Decay length scales are given in Table \ref{tab:decay}. We find that $\ell^+$ grows with larger texture size $\lambda^+$. That is, larger texture sizes show more nonlocal effects. This likely contributes to why the homogenized slip length model performs poorly for $\lambda^+ > \mathcal{O}(10)$ \cite{seo2016scaling}.

In the limit of small pattern size, the kernel is purely local, so we expect $\ell^+/\lambda^+ = 0$. In the limit of large pattern size, $\ell^+$ is limited by the channel half height and $\lambda^+$ is very large, so we again expect $\ell^+/\lambda^+ = 0$. Thus, there is some form of nonmonotonic behavior in the ratio of $\ell^+/\lambda^+$ as $\lambda^+$ increases. In the square post cases, $\ell^+/\lambda^+ = 0.06$ for $\lambda^+ = 77.6$,  $\ell^+/\lambda^+ = 0.08$ for $\lambda^+ = 103.4$, and $\ell^+/\lambda^+ = 0.09$ for $\lambda^+ = 155.1$. For the streamwise ridge case, $\ell^+/\lambda^+ = 0.09$.

\begin{table}[!htbp]
\caption{Decay length scale of slip length kernels.\label{tab:decay}}
\begin{ruledtabular}
\begin{tabular}{|c|c|c|c|c|c|}
  Case & P077 & P103 & P155 & R155 & S155 \\
  \hline
  $\lambda^+$ & 77.6 & 103.4 & 155.1 & 155.1 & 155.1 \\
  \hline
  $\ell^+$ & 5.2 & 8.7 & 13.6 & 14.2 & 20.6 \\
  \hline
  $\ell^+/\lambda^+$ & 0.06 & 0.08 & 0.09 & 0.09 & 0.13 \\
\end{tabular}
\end{ruledtabular}
\end{table}

The nonlocality percentages of the momentum transport cases can be found in Table \ref{tab:nonlocal}. Again, we find that larger texture size $\lambda^+$ corresponds to more significant effects arising from nonlocality. The nonlocality percentage is noticeably affected by the difference between pattern geometries of square posts and streamwise-aligned ridges. This behavior is perhaps predictable from close examination of Figure \ref{fig:slip_kernel}. The kernel associated with streamwise-aligned ridges has a distinctly different shape from those of the square posts, especially near the wall. The decay length scale, which is measured from kernel contributions further from the wall, seems to depend more upon $\lambda^+$ than on geometry. One can interpret decay length scale as the extent of nonlocality and nonlocality percentage as the magnitude of nonlocality. Comparing the post and ridge SHS of the same pattern size ($\lambda^+=155.1$), we observe that the extent of nonlocality does not differ; however, the magnitude of nonlocality in the ridge case is noticeably larger.

\begin{table}[!htbp]
\caption{Integral representation of nonlocality of slip length kernels. The second row represents the local contributions, the third row represents the nonlocal contributions, and the fourth row is the sum of both local and nonlocal contributions. The final row is the percentage of nonlocal compared to total contributions.\label{tab:nonlocal}}
\begin{ruledtabular}
\begin{tabular}{|c|c|c|c|c|c|}
  Case & P077 & P103 & P155 & R155 & S155 \\
  \hline
  $\lambda^+$ & 77.6 & 103.4 & 155.1 & 155.1 & 155.1 \\
  \hline
  $\int_0^{0^+} \beta^+ dy_2$ & 14.97 & 17.06 & 20.32 & 13.33 & 22.16 \\
  \hline
  $\int_{0^+}^{2h} \beta^+ dy_2$ & 1.04 & 2.61 & 4.23 & 4.00 & 9.01 \\
  \hline
  $\int_0^{2h} \beta^+dy_2$ & 16.01 & 19.50 & 24.55 & 17.33 & 31.17 \\
  \hline
  Nonlocal \%  & 6\% & 13\% & 17\% & 23\% & 29\%\\
\end{tabular}
\end{ruledtabular}
\end{table}

\subsection{Comparison of scalar and momentum transport}

\begin{figure}[!htbp]
  \centering
  \includegraphics{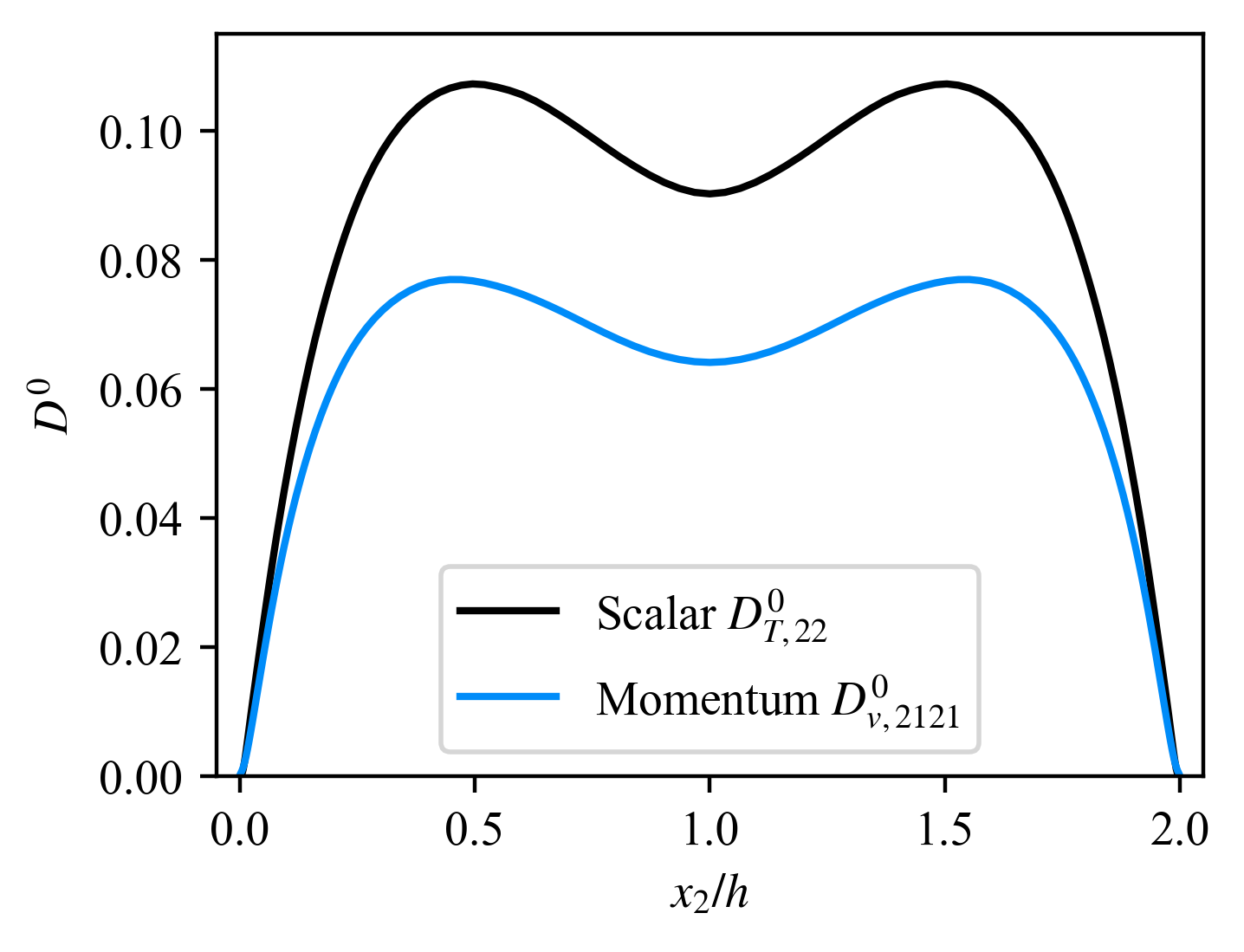}
  \caption{Eddy diffusivity (green) and eddy viscosity (blue) from scalar and momentum transport, respectively.}
  \label{fig:comp_eddy}
\end{figure}

\begin{figure}[!htbp]
  \centering
  \includegraphics{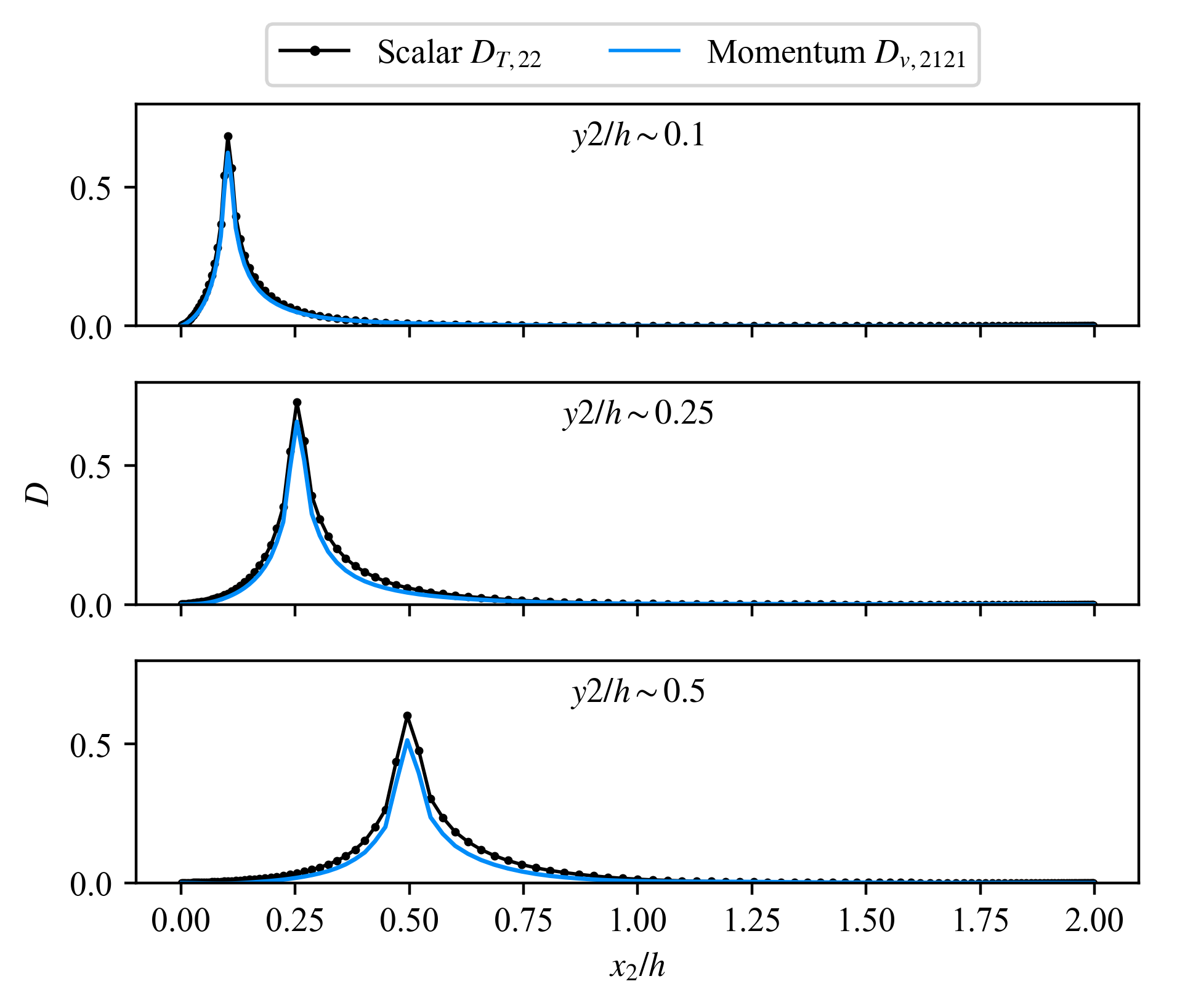}
  \caption{Comparison of eddy diffusivity (green) and eddy viscosity (blue) profiles of scalar and momentum transport for fixed $y_2$}
  \label{fig:comp_Dcuts}
\end{figure}

We now present a direct comparison of our results from assessing scalar and momentum transport of the $\lambda^+ = 155.1$ square post case. The leading order eddy diffusivity from scalar transport and the leading order eddy viscosity from momentum transport are shown in Figure \ref{fig:comp_eddy}. The nonlocal eddy diffusivity and eddy viscosity kernels are compared in Figure \ref{fig:comp_Dcuts}. The eddy viscosity of momentum transport has a narrower band of nonlocality relative to the eddy diffusivity of scalar transport.

\begin{figure}[!htbp]
  \centering
  \includegraphics{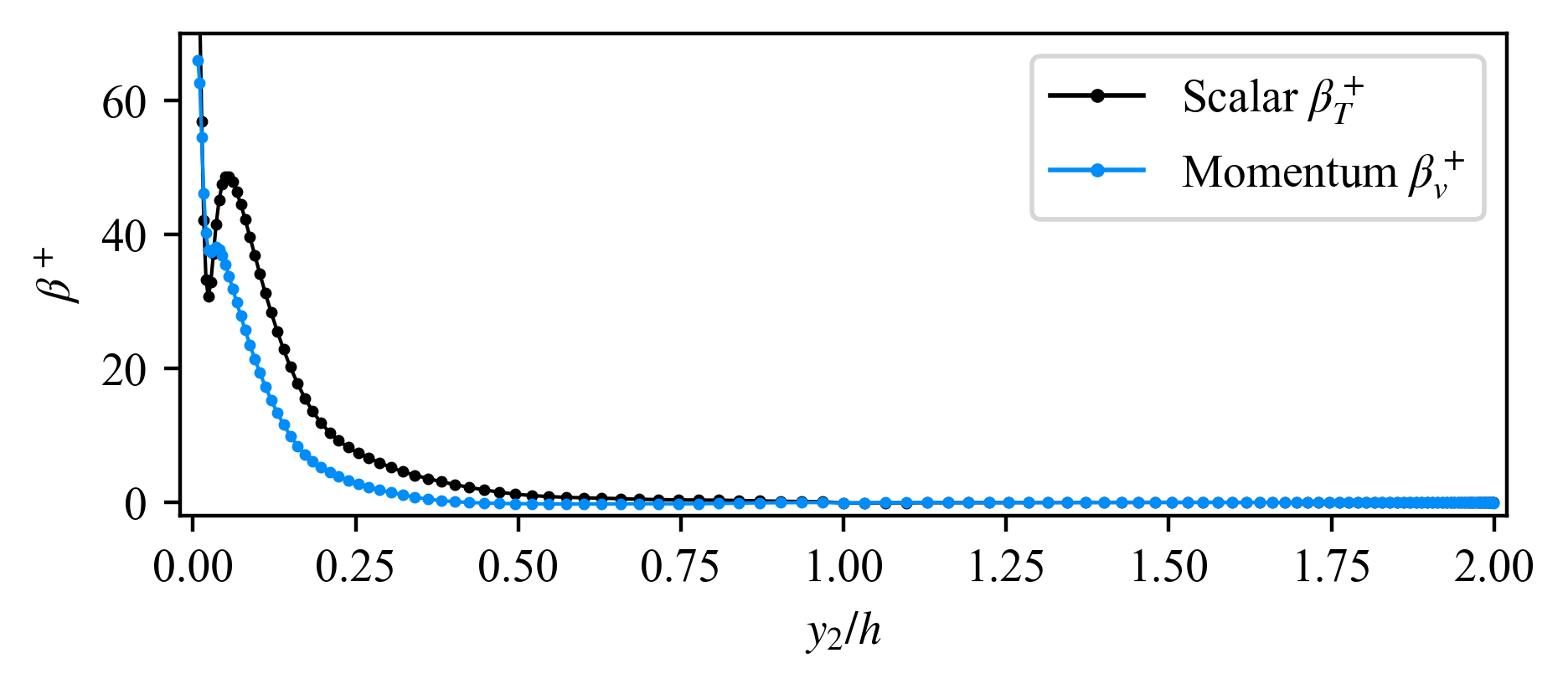}
  \caption{Nonlocal slip kernel of scalar transport (green) and momentum transport (blue) for pattern geometry $\lambda^+ = 155.1$}
  \label{fig:comp_slip}
\end{figure}

Figure \ref{fig:comp_slip} shows the nonlocal slip kernels of scalar and momentum transport. The slip kernel associated with scalar transport is visibly more nonlocal, as compared to the kernel associated with momentum transport. Decay length scales and nonlocality percentage are given in Tables \ref{tab:decay} and \ref{tab:nonlocal}, and confirm this observation: the decay length scale and nonlocality percentage of the scalar result are both significantly larger in magnitude than the momentum result.

This may, at first glance, be an unintuitive result. The scalar transport equation, as discretized in our numerical solver, is a local equation; that is, all terms in the equation are calculated with finite differences involving only neighboring points. The momentum transport equation, on the other hand, is nonlocal in the sense that the pressure gradient term is calculated by considering not just neighboring points, but all points in the computational domain. However, this finding is consistent with previous MFM analysis of homogeneous isotropic turbulence \cite{shirian2022eddy}. Recall that in our homogenized setting we have averaged in the streamwise and spanwise directions, as well as in time. Thus, advection has also become a nonlocal effect in this framework. Consistent with Shirian and Mani (2022) \cite{shirian2022eddy}, we conclude that the nonlocal advection effects are more prominent in scalar transport, or that nonlocal advection and pressure may partially cancel in momentum transport.

\subsection{RANS assessment}

We now use the local and nonlocal eddy viscosity and slip length operators to calculate the solution to the RANS equations (Eq. \ref{eqn:RANS}) for pattern size $\lambda^+ = 155.1$. Solutions were calculated with leading order eddy viscosity $D_{v,2121}^0$ and local slip length $\beta_v^0$, leading order eddy viscosity $D_{v,2121}^0$ and nonlocal slip length $\beta_v(y_2)$, and nonlocal eddy viscosity $D_{v,2121}(x_2,y_2)$ and nonlocal slip length $\beta_v(y_2)$. 

Using leading order eddy viscosity and local slip length results in the largest error between DNS and RANS solutions. The maximum error is a difference of 4.5 at the boundaries. The incorporation of nonlocal slip length kernel, while still using leading order eddy viscosity, greatly improves performance in the near wall region. The maximum error is 1.1 and the location is shifted to the buffer layer ($x_2^+ \approx 20$). The nonlocal eddy viscosity operator further increases the accuracy of the RANS calculation; the maximum error is 0.6. The RMS errors are 3.5 for leading order eddy viscosity and local slip length, 0.9 for leading order eddy viscosity and nonlocal slip length, and 0.2 for nonlocal eddy viscosity and nonlocal slip length. We see the greatest improvement in the accuracy of the RANS solution with the inclusion of both nonlocal slip and nonlocal eddy viscosity.

The improvement in the RANS solutions due to the inclusion of nonlocality in the eddy viscosity and slip length operators is particularly evident for the $\lambda^+ = 155.1$ case, as we have previously measured the decay length scale and nonlocality percentage to be significant. For the smaller pattern sizes of $\lambda^+ = 77.6$ and $\lambda^+ = 103.4$, the measured influence of using the nonlocal operators is on the order of numerical and statistical error.

\begin{figure}[!htbp]
  \centering
  \includegraphics{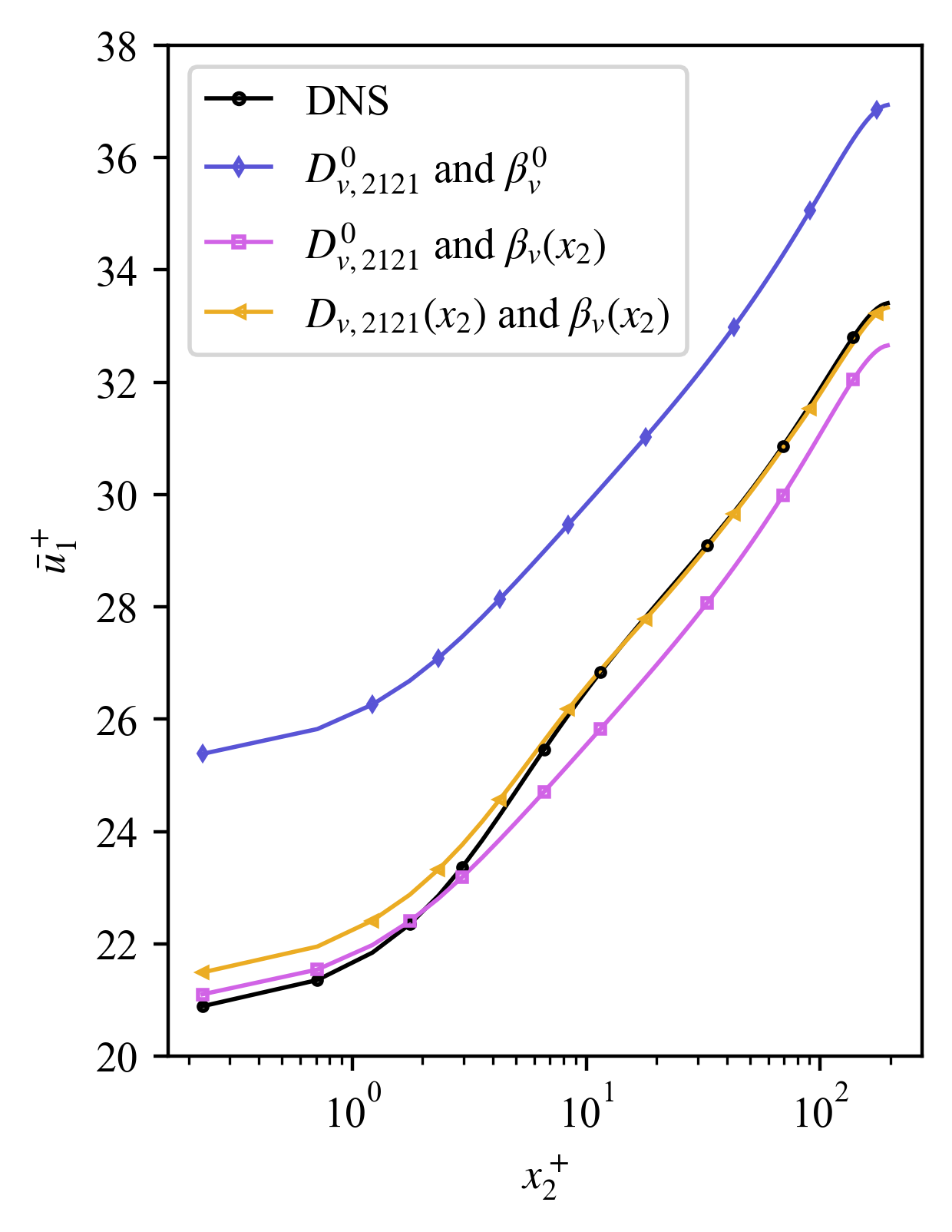}
  \caption{Mean velocity profiles of DNS (black), RANS with $D_{v,2121}^0$ and $\beta_v^0$ (purple), RANS with $D_{v,2121}^0$ and $\beta_v(x_2)$ (pink), and RANS with $D_{v,2121}(x_2)$ and $\beta_v(x_2)$ (yellow) for pattern size $\lambda^+ = 155.1$.}
  \label{fig:rans_155}
\end{figure}

\section{\label{sec:concl}Conclusions}
We have simulated superhydrophobic surfaces in turbulent channel flow with both pattern-resolved and homogenized slip length boundary conditions. Using the macroscopic forcing method, we have calculated nonlocal eddy diffusivity and slip length operators for streamwise ridges and square posts of texture wavelengths $\lambda^+ \approx 77-155$. 

Consistent with findings from Seo and Mani (2016) \cite{seo2016scaling}, we find disagreement in the mean behavior between corresponding pattern-resolved and homogenized slip length simulations. In probing the root cause of these differences, we have found significant differences in the momentum mixing behavior of these simulations, indicating that use of a homogenized slip length model may not be well-founded, even for small $\lambda^+$. From our calculations of the slip length kernel, we show that the effect of patterning on slip velocity at the wall is nonlocal for finite $Re$. We present two metrics for quantifying the relative importance of the nonlocal velocity gradients on the behavior at the wall. The length scale $\ell^+$ is defined to be the rate of decay of an exponential fit to the nonlocal slip kernel, and quantifies the extent of nonlocality. We also compare the integrated nonlocal components of the kernel to the leading order slip length, quantifying the magnitude of the nonlocality. Solutions to the RANS equations reveal that usage of nonlocal eddy viscosity and slip length operators shows significant improvement compared to leading order operators, indicating that the model form of a slip length boundary condition for patterned SHS should be treated as a nonlocal operator.

We have additionally calculated eddy diffusivity and slip length operators for scalar transport on square posts of texture wavelength $\lambda^+ \approx 155$. In our analysis of scalar and momentum transport on this superhydrophobic surface geometry, we find that scalar transport is more nonlocal than momentum transport in a homogenized setting. This is consistent with previous findings utilizing the macroscopic forcing method \cite{shirian2022eddy}.

In this work, we have assumed flat interface and zero Weber number; however, in real scenarios, the fluid-water interface will deform under a mean flow \cite{ling2016high,gose2021turbulent,fu2019experimental}. We expect the slip length to remain nonlocal as long as the superhydrophobic surface remains non-wetted. However, formation of bulges and troughs, as well as potential unsteadiness, will likely affect the shape, magnitude, and extent of the nonlocality kernels. The superhydrophobic surfaces will also not be regularly patterned, but will instead have random textures \cite{alame2019wall,seo2018effect}. Again, we expect the slip kernel associated with the random textures to be nonlocal, but we cannot at this point quantify the changes in the kernel due to texture randomness without additional DNS studies. Additionally, in real scenarios, the $Re$ will be significantly higher than in our simulations here. Below the $Re$ threshold for robustness of gas bubbles \cite{seo2018effect,wexler2015shear,srinivasan2015sustainable,ling2017effect}, $\lambda^+$ scales linearly with higher $Re$, and therefore the nonlocal effects are expected to be more important as $Re$ increases.

Finally, we would like to emphasize that this work's focus was on diagnosing the nonlocality of the slip mechanism and momentum mixing, and our results should not be interpreted as a standalone closure model. The quantified nonlocality in slip and eddy diffusivity behavior is expected to be geometry dependent, and the test cases presented here do not cover a comprehensive dataset from which we can interpret closure models for either RANS or texture-homogenized simulations. Given that the tools used here have, in other contexts, demonstrated promising potential for closure modeling \cite{shende2024model,liu2023systematic}, we envision the present findings as an important first step towards future modeling efforts.

\FloatBarrier

\begin{acknowledgments}
  This work was supported by the Office of Naval Research (ONR) under Grant N00014-22-1-2323.
\end{acknowledgments}

\newpage

%

\end{document}